\documentclass[twocolumn,amsmath,amssymb]{revtex4}
\usepackage{graphicx}
\usepackage{dcolumn}
\usepackage{bm}

\begin{document}

\title{Decay of highly-correlated spin states in a dipolar-coupled solid}\

\author{HyungJoon Cho\footnote{Currently address: Schlumberger Doll Research, Ridgefield, CT}, Paola Cappellaro, David G. Cory and Chandrasekhar Ramanathan\footnote{Author to whom correspondence should be addressed. Electronic address:sekhar@mit.edu}}
\affiliation{Department of Nuclear Science and Engineering, Massachusetts Institute of Technology, Cambridge 02139, USA}

\date{\today}

\begin{abstract}
We have measured the decay of NMR multiple quantum coherence intensities both under the internal dipolar Hamiltonian as well as when this interaction is effectively averaged to zero, in the cubic calcium fluoride (CaF$_2$) spin system and the pseudo one-dimensional system of fluoroapatite.  In calcium fluoride the decay rates depend both on the number of correlated spins in the cluster, as well as on the coherence number.  For smaller clusters, the decays depend strongly on coherence number, but this dependence weakens as the size of the cluster increases.  The same scaling was observed when the coherence distribution was measured in both the usual Zeeman or $z$ basis and the $x$ basis.  The coherence decay in the one dimensional fluoroapatite system did not change significantly as a function of the multiple quantum growth time, in contrast to the calcium fluoride case.  While the growth of coherence orders is severely restricted in this case, the number of correlated spins should continue to grow, albeit more slowly.  All coherence intensities were observed to decay as  Gaussian functions in time.  In all cases the standard deviation of the observed decay appeared to scale linearly with coherence number.

\end{abstract}


\maketitle
\section{Introduction}
While there have been several proposals put forward for scalable quantum computing architectures, experimental realizations have been limited to a handful of qubits at most.  Maintaining coherence as the size of the system Hilbert space increases remains extremely challenging.  It is essential to understand how decoherence rates in different physical systems scale as a function of system size.  
There have been a number of theoretical investigations on the scaling behaviour of decoherence \cite{Unruh,Palma,Duan,Reina}.  These general models have typically been based on the spin-boson model of Leggett and co-workers \cite{Leggett}.  System-specific scaling laws have also been proposed for a few physical implementations (e.g. \cite{Dalton,Ischi}).

Palma {\em et al}.\ \cite{Palma} showed that for a multi-qubit quantum register, the decay of particular off-diagonal elements of the system density matrix depended on the Hamming distance $f$ between the two states.  In the case of independent, uncorrelated noise, the decay was of the form $\exp(-f\Gamma(t))$ while for the case of correlated noise the decay was $\exp(-f^2\Gamma(t))$ where $\Gamma(t)$ corresponds to the single qubit decay. 

Suter and coworkers recently published an experimental study of the decay of multi-spin states using nuclear magnetic resonance (NMR)\cite{Krojanski}.  They used multiple-quantum (MQ) NMR 
experiments to create correlated multi-spin states in a powdered sample of the plastic crystal adamantane, and observed the rate at which these states decay during evolution under the internal dipolar coupling of the spins.  They observed that the decay rate increased as a square root of the estimated number of correlated spins in the cluster.   

A theoretical analysis of their experimental results has been published recently by Fedorov and Fedichkin \cite{Fedorov}.  Neglecting the flip-flop (XY) terms of the dipolar interaction, they  obtained the following expression for the decay of multiple quantum coherence states, in the short time, large spin limit 
\begin{equation}
S_n(t) = p\exp\left(-\alpha n^2t^2\right) + \left(1-p\right)\exp\left(-\frac{N}{2}\alpha t^2\right)
\label{eq:Fedorov}
\end{equation}
where $p = 1/N (\sum_j d_{jk})^2 / \sum_j d_{jk}^2$ is a correlation parameter, $\alpha$ is proportional to the second moment of the lineshape, $n$ is the coherence number and $N$ is the number of correlated spins.  The first term depends strongly on the coherence number and indicates that the spins in the multi-quantum state experience a correlated mean field.  The second term does not depend on the coherence number, but only depends on the number of spins in the cluster, and indicates that the fields experienced by the different spins are uncorrelated from each other.  

These equations agree well with the measured data at short times, but deviate at longer times.  The correlation parameter $p$ appeared to remain constant with increasing spin system size.  However since adamantane is a plastic crystal in which the molecules are undergoing rapid rotational motions, the intramolecular dipolar couplings are averaged to zero, and the only residual couplings are motionally averaged intermolecular dipolar couplings.  This motional averaging modulates the contributions in Equation 1, making the experimental results obtained peculiar to this class of samples.  

In this paper we expand on these preliminary studies.  Our test systems are the cubic lattice of 100\% abundant $^{19}$F spins in a single crystal of calcium fluoride, and the pseudo one dimensional spin chains of fluoroapatite (FAp).  These are both rigid crystals, and molecular motions will not affect the results obtained.  In addition to characterizing the decay of the system under the dipolar Hamiltonian, we measure the decay rates obtained when we suspend evolution under the dipolar Hamiltonian.  We also repeated the experiments, encoding the multi-spin coherences in a different basis, and observed the resulting scaling behaviour.  

The standard MQ NMR experiment is shown in Figure 1 
\cite{Baum,Munowitz}.  Assuming that the system is closed, the final density matrix is given by 
\begin{equation}
\rho_f^{\zeta}(\phi) = U_{DQ}^\dag U_{ev}U_{\zeta}(\phi) U_{DQ} \rho_i U_{DQ}^{\dag} U_{\zeta}^\dag(\phi) U_{ev}^\dag U_{DQ}
\label{eq:unitaries}
\end{equation}
$\phi$ is the phase angle and $\zeta = \{x,z\}$ is the rotation axis for the multiple-quantum encoding.  Collective rotations about each axis yield coherence order distributions in the corresponding basis.  The propagator $U_{DQ} = \exp\left(-it\mathcal{H}_{DQ}\right)$ represents evolution under the double quantum (DQ) Hamiltonian given by 
\begin{equation}
\mathcal{H}_\mathrm{DQ}=-\frac{1}{2}\sum_{j<k}D_{jk}\left\{\sigma_{j}^{+}\sigma_{k}^{+}+\sigma_{j}^{-}\sigma_{k}^{-}\right\}.\\
\end{equation}
where the dipolar coupling constant $D_{jk}$ between spins $j$ and $k$ is
\begin{equation}
D_{jk}=\frac{\gamma^{2}\hbar^{2}}{{r}_{jk}^{3}}\left(1-3\cos^{2}\theta_{jk}\right),\\
\end{equation}
$\gamma$ is the gyromagnetic ratio, $r_{jk}$ is the distance
between spins $j$ and $k$, and $\theta_{jk}$ is the angle between
the external magnetic field and inter-nuclear vector $\vec{r}_{jk}$.  This effective DQ Hamiltonian is created by multiple-pulse NMR techniques that toggle the dipolar Hamiltonian to create the appropriate zeroth order Average Hamiltonian.
In Equation~\ref{eq:unitaries} above we have assumed that the experimental implementation of $-\mathcal{H}_{DQ}$ is perfect.
The encoding of the coherence orders is performed by the collective rotations $U_\zeta(\phi) = \exp\left(-i\phi\sum_j \sigma_{\zeta j}\right)$.
The observed signal is given by the overlap
\begin{eqnarray}
S(t) & = & Tr\left[\rho_f^\zeta (\phi)\rho_i \right] \\
& = & Tr\left[\rho_{f}^{DQ\zeta} (\phi) \rho_{i}^{DQ}\right]
\end{eqnarray}
where  $\rho_{f}^{DQ\zeta}(\phi) = U_{ev} U_\zeta(\phi) \rho_{iDQ} U_\zeta^\dag (\phi) U_{ev}^\dag$, and $\rho_{i}^{DQ} = U_{DQ}\rho_i U_{DQ}^\dag$, and the evolution $U_{ev}$ is defined below.
 
Following this evolution the density operator of the spin system in the Zeeman basis  contains off-diagonal terms of the form
\begin{eqnarray}
 \sigma_+^1\sigma_+^2\cdots \sigma_+^n\sigma-^{n+1}\cdots \sigma_-^{n+m}\sigma_z^{n+m+1}\cdots \sigma_z^{n+m+q} \nonumber \\ +\hspace{0.02in} \sigma_-^1\sigma_-^2\cdots \sigma_-^n\sigma_+^{n+1}\cdots \sigma_+^{n+m}\sigma_z^{n+m+1}\cdots \sigma_z^{n+m+q}. 
\end{eqnarray}
We are interested in the properties of such coherences in the system.  In the experiments here, we create a distribution of states with different coherence orders $M = (m-n)$, and spin numbers $K = n+m+p$.  
After this step, we either allow the spin system to evolve under the internal dipolar
interaction or suspend evolution of the spin system by applying a  time suspension sequence. In the first case, $U_{ev} = \exp\left(-i\mathcal{H}_D t\right)$, where the dipolar
Hamiltonian is
\begin{equation}
\mathcal{H}_D=\sum_{j<k}D_{jk}\left\{\sigma_{jz}\sigma_{kz}-\frac{1}{4}(\sigma_{j+}\sigma_{k-}+\sigma_{j-}\sigma_{k+})\right\}.
\label{eq:dipH}
\end{equation}
In the second case $U_{ev} = \mathcal{I}$, if the time suspension sequence is perfect and the system is completely isolated.  Decays observed during this time could be the result of errors in the control, or couplings to an environment.

\begin{figure}
\scalebox{0.5}{\includegraphics{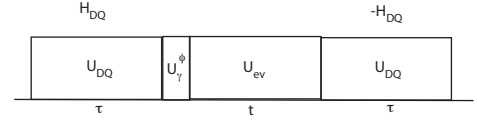}}
\caption{\label{pulsesequencecontrol} A generic MQ NMR experiment.  A $\pi/2$ pulse is applied at the end of the pulse sequence for 
signal acquisition.}
\end{figure}

\begin{figure}
\scalebox{0.42}{\includegraphics{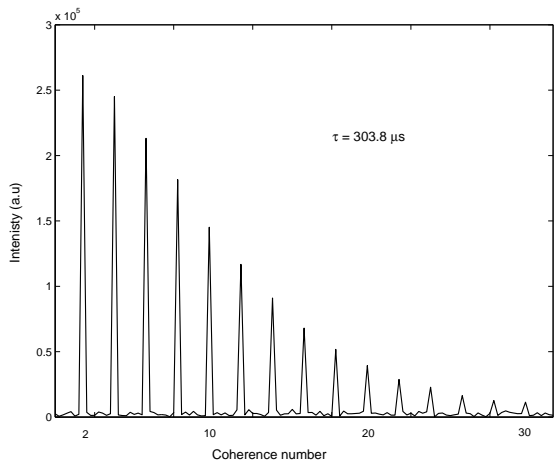}}
\caption{\label{coherenceloop7}Zeeman basis coherence order distribution in CaF$_2$ with
$t$=0, for an MQ growth time $\tau$=303.8 $\mu$s}
\end{figure}

 The experiments were
performed at room temperature at 2.35 T (94.2 MHz, $^{19}$F), using
a Bruker Avance spectrometer and home built probe. The samples used
were a 1 mm$^{3}$ single crystal of CaF$_{2}$ with T$_{1}$ $\sim$ 7s, and a crystal of fluorapatite (FAp) with $T_1 \sim  200$ ms. The FAp crystal is a mineral crystal specimen from Durango, Mexico.
High power 0.5 $\mu s$ $\pi$/2 pulses were used.  The rotation $U_z(\phi) = \exp\left(-i\phi\sum_j \sigma_{zj}\right)$ was used to encode coherence number in the Zeeman or $z$ basis while the rotation $U_x(\phi) = \exp\left(-i\phi\sum_j \sigma_{xj}\right)$ was used to encode coherence orders in the $x$-basis \cite{Ramanathan,vanBeek}.  The phase ($\phi$) was incremented from 0 to 4$\pi$ with $\Delta\phi=\frac{\pi}{32}$ to encode up to 32 quantum coherences for every experiment. A
fixed-time point corresponding to the maximum intensity signal was
sampled for each $\phi$ value, and then was Fourier transformed with
respect to $\phi$ to obtain the coherence order distribution, as seen in Figure~\ref{coherenceloop7}.  

As the evolution time $\tau$ under the DQ Hamiltonian increases (also referred to as the MQ growth time), progressively more spins are correlated into the MQ states.  Table 1 shows the size of the spin system for each MQ growth time, as estimated by the method of Baum {et al}.\ \cite{Baum}, and used in \cite{Krojanski}.  A log-log fit indicates that the number of spins is increasing as $N (= n+m+q) \sim t^2$.

\begin{center}
\begin{tabular}{|c|c|} \hline
{\small MQ growth time} & {\small System size (N)} \\ \hline
{\small 43.3 $\mu$s} & {\small 6} \\
{\small 86.8 $\mu$s} & {\small 8} \\
{\small 130.2 $\mu$s} & {\small 12} \\
{\small 173.6 $\mu$s} & {\small 28} \\
{\small 217 $\mu$s} & {\small 36} \\
{\small 260.4 $\mu$s} & {\small 66} \\
{\small 303.8 $\mu$s} & {\small 96} \\ \hline 
\end{tabular}

{\small Table 1.  Effective size of the spin cluster as a function of the MQ growth time, using the model in \cite{Baum}}.
\end{center}

\section{Evolution under the secular dipolar Hamiltonian}
\begin{figure}
\scalebox{0.45}{\includegraphics{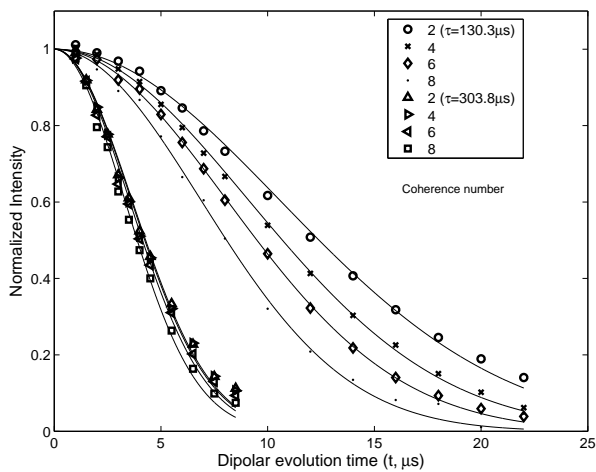}}
\caption{\label{dipolardecay}Decay of intensity for the different coherence orders for MQ growth times $\tau=130.3 \mu$s, and $\tau=303.8 \mu$s in CaF$_2$. The solid lines are Gaussian fits to
the data. }
\end{figure}

The MQ states are not stationary under the internal dipolar Hamiltonian of the spins, and evolve as a function of the dipolar interaction time.  This uncompensated evolution leads to imperfect refocusing during $-\mathcal{H}_{DQ}$.
Figure~\ref{dipolardecay} shows the intensities of various coherence
orders in the $z$ basis as a function of dipolar evolution time for two different MQ growth times.  The signal intensity appears to decay as a Gaussian function in time.  We have fit the data to $\exp(-t^2/2T_d^2)$ to extract the effective decay times ($T_d$ = standard deviation) for each coherence order.  The Gaussian shape of the decays indicates that the underlying process is consistent with a {\em time-invariant} dispersion of fields (due to the spins) having a normal distribution \cite{Abragam}.

\begin{figure}
\scalebox{0.45}{\includegraphics{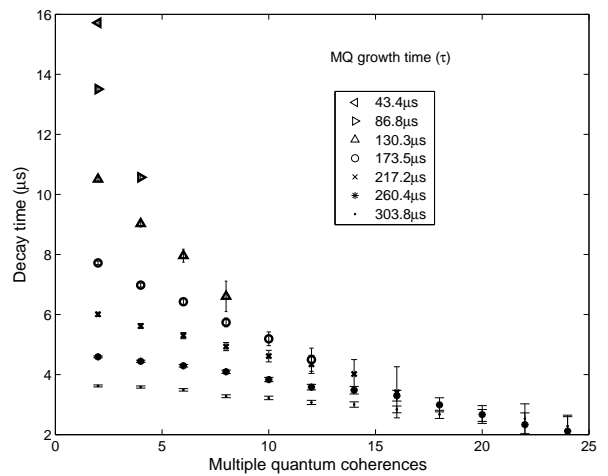}}
\caption{\label{decaydipolar}Effective decay times under the dipolar Hamiltonian of the various
coherence orders for different MQ growth times in CaF$_2$.}
\end{figure}

Figure~\ref{decaydipolar} shows the decay times $T_d$ of the different $z$ basis coherence orders as a function of the MQ growth times.  Four features are evident in the data:  (i) the $T_d$ are seen to depend linearly on the coherence number;  (ii) for short $\tau$, the $T_d$ are seen to depend strongly on the coherence number; (iii) this dependence on coherence number weakens significantly with increasing $\tau$ and (iv) the incremental change in $T_d$ with increasing $\tau$ decreases.

\begin{figure}
\scalebox{0.45}{\includegraphics{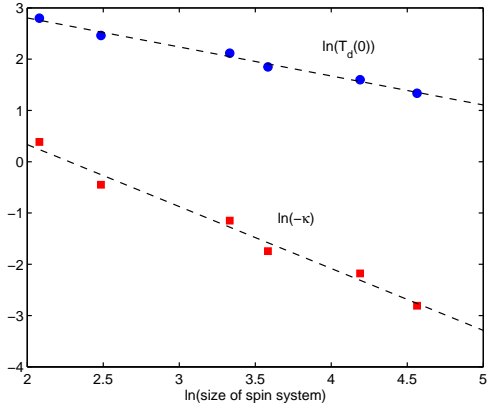}}
\caption{\label{fig:intercept} Dependence of $\ln(T_d(0))$ and $\ln(-\kappa)$ on $\ln(N)$ the size the spin system, as well as the best linear fits obtained.}
\end{figure}

We performed a linear fit of $T_d$ versus coherence number for each of the MQ growth times as 
\begin{equation}
T_d = T_d(0) - \kappa \cdot n
\end{equation}
where $T_d(0)$ is the decay of the zero-quantum intercept, and $\kappa$ is the slope.  Figure~\ref{fig:intercept} shows the dependence of $\ln(T_d(0))$ and $\ln(-\kappa)$ on $\ln(N)$ the size of the spin system from Table 1, as well as the best linear fits obtained. We get
$\ln(T_d(0)) = 3.94 - 0.57\cdot\ln(N)$ and $\ln(-\kappa) = 2.75 - 1.21\cdot\ln(N)$.  Thus $T_d(0) \approx A/\sqrt{N}$ and $\kappa \approx -B/N$, where $A=51.4$ and $B=15.6$.  We can therefore express the scaling behavior of $T_d$ as 
\begin{equation}
T_d = \frac{A}{\sqrt{N}} - \frac{B\cdot n}{N} \: \: .
\end{equation}

\begin{figure}
\scalebox{0.45}{\includegraphics{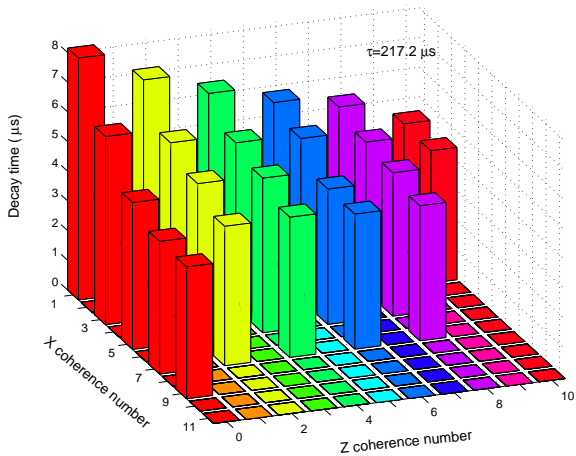}}
\caption{\label{decaydipolar3d}Effective decay times under the dipolar
Hamiltonian for correlated
$x$ and $z$ bases coherence orders in CaF$_2$.}
\end{figure}

We have repeated the experiment encoding the coherence orders in  the $x$-basis instead, and obtained identical scaling behaviour.   The decays were once again observed to be Gaussian.  In Figure~\ref{decaydipolar3d} we show the results of a multi-dimensional experiment in which we correlate the $z$ and $x$ basis decay times for an MQ growth time $\tau =$ 217.2 $\mu$s.  It should be noted that the dipolar Hamiltonian in a strong external magnetic field is anisotropic (see Equation~\ref{eq:dipH}), but this is not reflected in the observed decay times.

\section{Evolution under a time suspension sequence}

We then attempted to suppress evolution of the dipolar Hamiltonian using a time-suspension pulse sequence that implements (approximately) the identity operator on the spin system \cite{Cory}.  In the ideal experiment, we should see no decay of the spin coherences due to dipolar couplings within the spin system.  The cycle time of the 48-pulse time suspension sequence  used here  was 132.48 $\mu s$.  The change in 
intensity of the $z$-basis coherence order was measured as a function of the number of loops of the 48-pulse sequence.

\begin{figure}
\scalebox{0.45}{\includegraphics{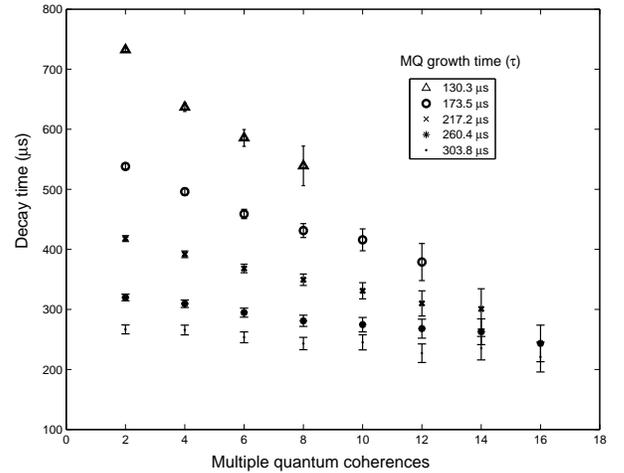}}
\caption{\label{decay48pulse}Effective decay times under the 48-pulse
sequence of the various
coherence orders for different MQ growth times in CaF$_2$ .}
\end{figure}

Once again we observed Gaussian decays as a function of time for the coherence intensities, indicating that the underlying noise has a long correlation time.  We fit the data to a Gaussian and extracted the $T_d$ decay times.
Figure~\ref{decay48pulse} shows the effective $T_d$'s of the $z$-basis coherence orders, under the time-suspension sequence.  We see that the  data is qualitatively identical to that obtained in the case of dipolar evolution, showing the same features discussed above.  This correspondence was repeated in the $x$-basis data, as well as in the correlated $z$ and $x$ basis encoding experiments.

\begin{figure}
\scalebox{0.4}{\includegraphics{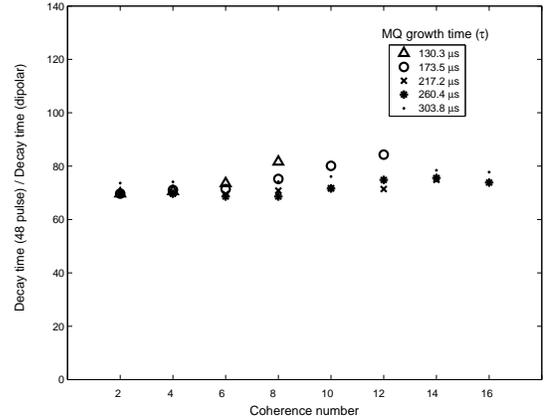}}
\caption{\label{ratio}Ratio of decay time under the 48-pulse
sequence to decay time under the dipolar evolution in CaF$_2$.}
\end{figure}

Figure~\ref{ratio} shows the ratio of the $z$-basis multiple quantum $T_d$'s during evolution under the 48-pulse time suspension sequence to the $T_d$' measured during dipolar evolution.   The ratio was measured to be around 70, and appeared to be independent of the size of the spin correlations involved.   Given the uniform scaling obtained, we did not repeat the linear fits.  However it is clear from the uniform scaling that both $A$ and $B$ are scaled by the same factor of around 70.

\section{quasi-1D Fluoroapatite system}
We also measured the decay of multiple quantum coherences under the dipolar Hamiltonian in fluoroapatite.  Floroapatite is a quasi-one dimensional spin chain, as the distance between spin chains is about three times larger than the distance between adjacent spins in the chain \cite{Engelsberg}.  The one dimensional spin chain with nearest neighbor double quantum Hamiltonian is exactly solvable \cite{Feldman1,Feldman2}, and it has been shown that starting from a thermal equilibrium state, only zero and double quantum coherences are produced, even as higher order multi-spin states are created.  The presence of higher order coherences indicates the importance of next-nearest neighbor and other distant couplings.  Consequently it has been observed that higher order coherences grow very slowly in this system \cite{Yesinowski}.

Consistent with the one dimensional nature of the spin system, the growth of the spin clusters is much slower in this case.  More importantly, we see that the character of the decays does not change as a function of the the MQ growth time.  A strong dependence on coherence number is observed for both short and long MQ growth times.
However, even at long times, the number of correlated spins is still small in this system.

\begin{figure}
\scalebox{0.45}{\includegraphics{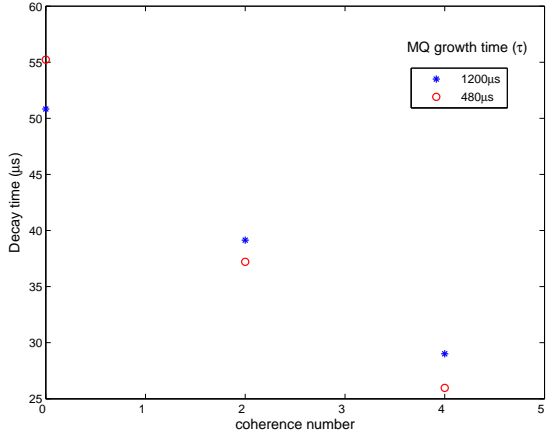}}
\caption{\label{FAPdip}Effective decay times under the dipolar Hamiltonian in fluoroapatite for MQ growth times $\tau = 480 \mu$s and $\tau = 1200 \mu$s}
\end{figure}
 
\section{Discussion}

It is important to ensure that the observed decays are not an artifact of the encoding used to make the measurement.  For example, we need to examine if the phases introduced into the system by the encoding $U_\zeta(\phi)$ lead to imperfect refocusing in our experiments.  We measured the decay time for the particular situation where $U_\zeta(\phi) = U_\zeta^\dag(\phi) = \mathcal{I}$, as a function of the MQ growth time.  The observed scaling for the case of dipolar evolution was $T_d \approx 22.7 / \sqrt{N}$ and $T_d \approx 2276 / \sqrt{N}$ for the 48-pulse time suspension sequence.  These times are slightly shorter than the corresponding values of $T_d(0)$ for the zero quantum coherence.  This is reasonable as we are measuring the aggregate decay of all the MQ states here, and suggests that the encoding step is not responsible for the decay rates observed.
   
The similarity between the dipolar evolution and the time suspension data suggests that the dominant source of noise in the time suspension data are residual dipolar coupling terms that are not effectively averaged out, as these would be expected to scale identically in the two cases.  However, it is worth examining an alternative model in which the decay is due to coupling to an external environment.  Fedorov {\em et al}.\ have also calculated the effect of a large,  bath coupled to the multiple quantum states, and get a similar solution to that obtained under an inter-spin Ising coupling.  Thus both theories - residual dipolar errors and the presence of external spins yield identical scaling behaviours and it is not possible {\em a priori} to distinguish between theses two models on the basis of the data here.  However we can examine the physics of the system under study to understand the origin of the decays.

In principle the environment that the $^{19}$F nuclear spins are coupled to could be lattice phonons or other spins---both electron and nuclear---that are present in the system.  In order to effectively relax the nuclear spins, the phonons would need to be resonant with the Larmor frequency of the spins (94.2 MHz in these experiments).
The Debye Temperature of calcium fluoride is 510 K, so it can be assumed that the spins are fixed in a rigid lattice in these room temperature measurements, and that phonons do not play a significant role in the relaxation of the spins.  In addition to sparse paramagnetic impurities (responsible for $T_1$ relaxation), we know that the CaF$_2$ system contains $^{43}$Ca which has spin 7/2 and is 0.13$\%$ abundant.  There could possibly be other spin defects in the crystal such as protons---albeit at much lower concentrations.

\subsubsection*{Paramagnetic impurities}
We also measured the decay rates under the dipolar evolution in a second crystal with a longer $T_1$ relaxation time $\approx$ 110 s, corresponding to a smaller concentration of paramagnetic impurities, and found no difference in the experimentally observed $T_d$'s.   In Mn-doped CaF$_2$ with a $T_1$ of 700 ms, Tse and Lowe estimated that the concentration of impurities ($N_e$) was approximately $5.6 \times 10^{17}$ cm$^{-3}$ \cite{Tse}, yielding an average impurity separation  of 15 nm.  Since $1/T_1 \propto N_e$, a 7 s $T_1$ corresponds to an impurity concentration of $5.6 \times 10^{16}$ cm$^{-3}$ and an average separation of 32.4 nm, while for $T_1 = 110$s, we get a concentration of $3.56 \times 10^{15}$ cm$^{-3}$ and a separation of 81.3 nm.  

The magnetic moment of a S=5/2 paramagnetic impurity is about 3500 times larger than that of the fluorine nucleus.  The dipolar coupling between such an impurity and a $^{19}$F nucleus becomes comparable to the strongest $^{19}$F--$^{19}$F coupling ($a$ = 2.73 \AA) at a distance of about 4 nm.  This corresponds to an interaction strength of about 15 kHz, and dipolar correlation time of about 66 $\mu$s.  This is approximately the size of the frozen core of fluorine spins around each impurity site, in which the ``flip-flop'' or XY terms of the dipolar Hamiltonian are suppressed.  The electron-nuclear interaction further drops to 10 \% of the inter-nuclear coupling at a distance of 8.9 nm.   Thus, assuming Mn impurities, for a $T_1 = 7$ s crystal, 84 \% of the nuclear spins experience a hyperfine field that is less than 1.5 kHz.  The correlation time of the hyperfine field seen by the bulk of the nuclear spins is much too long to explain the observed decays.

\subsubsection*{$^{43}$Ca spins}
The magnetic moment of the calcium spins is about half that of the fluorine spins, and the Ca--F spacing is about 2.36 \AA, just a little shorter than the F--F distance.  Thus we expect the strongest Ca--F coupling to be 11.6 kHz.  However, given the low natural abundance of the Ca spins (0.13 \%), very few nuclear spins see a Ca coupling of this strength.  Another possibility is that the Ca--Ca dipolar coupling leads to mutual spin flips, that reduces the efficiency of the 48-pulse time suspension sequence.  The mean separation between $^{43}$Ca spins is 4.1 nm, and the average $^{43}$Ca--$^{43}$Ca coupling is about 2.2 Hz.  Since the Ca--F coupling is decoupled on the time-scale of one cycle of the 48-pulse sequence, which is about 44 $\mu$s, the 2 Hz Ca--Ca coupling is too weak to affect the efficiency of the decoupling sequence.  Thus we do not expect the $^{43}$Ca spins to be the source of the observed decays.

\subsubsection*{Errors in control}

We have assumed to this point that the implementation of all the pulse-sequences has been perfect.  The propagators $U_{DQ}$ and $U_{DQ}^\dag$ are idealized propagators for the NMR pulse sequences used, and typically correspond to the zeroth order Average Hamiltonian of the sequence.  In reality the presence of higher order terms in the Magnus expansion result in $U_{DQ}^{exp}U_{DQ}^{\dag exp} \ne \mathcal{I}$.  More importantly, errors in the implementation of the two sequences---the presence of phase transients during the leading and falling edges of the pulses, as well as errors in the setting of the $\pi/2$ pulse lead to imperfections in the refocusing.  This imperfect refocusing is however unlikely to be the source of the observed decay in either experiment.  In the dipolar evolution experiment, the strength of the dipolar coupling is much stronger than any residual error terms in the propagator, and it is these couplings that determine the decay rate.

In the time-suspension experiments, the 48-pulse sequence averages the dipolar interaction to zero to second order in the Magnus expansion.  Assuming perfect implementation, the leading error terms are likely to be the second-order dipolar-offset term and the second order offset term.  Our results indicate that the source of the observed signal decay in the time suspension experiments is the residual errors in the zeroth-order Average Hamiltonian, which are proportional to $\mathcal{H}_D$, rather than second-order terms like $\left[\mathcal{H}_D(t_1),\mathcal{H}_D(t_2)\right]$, or other higher-order terms.  The configurational space accessible by a single spin-flip is much smaller than that accessible by two spin-flips, especially at larger spin numbers.  This suggests that we should expect significantly different scaling of decay rates for the two types of processes.  Given the identical scaling behaviour observed in the two cases, we can conclude that the 
the dominant error in the time-suspension sequences has the form of $\mathcal{H}_D$ (possibly toggled about some arbitrary axis).  The more likely source of error in the control is imperfect implementation of the sequence.  Phase transients during the  pulses, or small errors in the setting of the $\pi/2$ pulse can accumulate over the hundreds of pulses that are applied in this experiment, and the residual error terms will look like modulated dipolar interactions, and consequently will scale the same way, though with a reduced strength \cite{Viola}.  

{\bf Acknowledgements}

This work was supported in part by the National Security Agency (NSA) under Army Research Office (ARO) contract numbers DAAD19-03-1-0125 and W911NF-05-1-0469, and by DARPA DSO and the Air Force Office of Scientific Research.

\bibliography{thebibliography}

\end{document}